\documentclass[sigconf]{acmart}

\acmConference[ICPC 2022]{The 30th International Conference on Program Comprehension}{May 21–22, 2022}{Pittsburgh, PA, USA}
\usepackage{caption}
\usepackage{subfigure}
\usepackage{hyperref}
\usepackage{multirow}
\usepackage{indentfirst}
\usepackage{appendix}
\usepackage{threeparttable}
\usepackage{tablefootnote}
\usepackage{bm}
\usepackage{graphicx}
\AtBeginDocument{%
  \providecommand\BibTeX{{%
    \normalfont B\kern-0.5em{\scshape i\kern-0.25em b}\kern-0.8em\TeX}}}

\copyrightyear{2022}
\acmYear{2022}
\setcopyright{acmcopyright}\acmConference[ICPC '22]{30th International Conference
on Program Comprehension}{May 16--17, 2022}{Virtual Event, USA}
\acmBooktitle{30th International Conference on Program Comprehension (ICPC '22),
May 16--17, 2022, Virtual Event, USA}
\acmPrice{15.00}
\acmDOI{10.1145/3524610.3527896}
\acmISBN{978-1-4503-9298-3/22/05}

\usepackage{xcolor}

\newcommand{\myauthornote}[3]{{\color{#2} {\sc #1}: #3}}
\newcommand\hy[1]{\myauthornote{HY}{red}{#1}}

\begin{document}

\bibliographystyle{unsrt}
\title{HELoC: Hierarchical Contrastive Learning of Source Code Representation}

\author{Xiao Wang}
\affiliation{%
 \institution{Shandong Normal University}
  \city{Jinan}
  \country{China}
  }
  
\author{Qiong Wu}
\affiliation{%
 \institution{Shandong Normal University}
  \city{Jinan}
  \country{China}
  }
  
\author{Hongyu Zhang}
\affiliation{%
 \institution{The University of Newcastle}
  \city{NSW}
  \country{Australia}
  }  
 
\author{Chen Lyu}
\authornote{Corresponding author. Email: lvchen@sdnu.edu.cn}
\affiliation{%
 \institution{Shandong Normal University}
  \city{Jinan}
  \country{China}
  }
 
\author{Xue Jiang}
\affiliation{%
 \institution{Shandong Normal University}
  \city{Jinan}
  \country{China}
  } 

\author{Zhuoran Zheng}
\affiliation{%
 \institution{Nanjing University of Science and Technology}
  \city{Nanjing}
  \country{China}
  } 
  
\author{Lei Lyu}
\affiliation{%
 \institution{Shandong Normal University}
  \city{Jinan}
  \country{China}
  }  
  
\author{Songlin Hu}
\affiliation{%
 \institution{Institute of Information Engineering, Chinese Academy of Sciences}
  \city{Beijing}
  \country{China}
  }  

\ccsdesc[500]{Software and its engineering~Software maintenance tools}

\renewcommand{\shortauthors}{Wang, et al.}

\begin{abstract}
\noindent Abstract syntax trees (ASTs) play a crucial role in source code representation. 
However, due to the large number of nodes in an AST and the typically deep AST hierarchy, it is  challenging to learn the hierarchical structure of an AST effectively. In this paper, we propose HELoC, a hierarchical contrastive learning model for source code representation. To effectively learn the AST hierarchy, we use contrastive learning to allow the network to 
predict the AST node level and learn the hierarchical relationships between nodes in a self-supervised manner, which makes the representation vectors of nodes with greater differences in AST levels farther apart in the embedding space. By using such vectors, the structural similarities between code snippets can be measured more precisely. In the learning process, a novel GNN (called Residual Self-attention Graph Neural Network, RSGNN) is designed, which enables HELoC to focus on embedding the local structure of an AST while capturing its overall structure.
HELoC is self-supervised and can be applied to many source code related downstream tasks such as code classification, code clone detection, and code clustering after pre-training. 
Our extensive experiments demonstrate that HELoC outperforms the state-of-the-art source code representation models.
\end{abstract}

\keywords{Abstract Syntax Tree, Contrastive Learning, Code Representation}


\maketitle

\section{Introduction}
\noindent In recent years, many deep learning based approaches to software engineering tasks, such as code retrieval \cite{gu2018deep,sachdev2018retrieval}, code clone detection  \cite{hua2020fcca,bui2021infercode,wang2020detecting,wang2020modular}, code comment generation  \cite{gros2020code,hu2018deep,wang2021automatic}, software defect prediction  \cite{dam2018automatic,pei2020amalnet,zhou2019devign}, and code summarization~\cite{ahmad2020transformer,zhang2020retrieval,wang2020reinforcement,leclair2020improved}, have emerged to improve software development and maintenance. These approaches require effective learning of the syntax and semantics of source code to obtain a proper neural code representation. 
In current research on source code representation, some approaches split the source code into tokens and encode the token sequence based on the CNN \cite{allamanis2016convolutional,huo2016learning} and RNN \cite{li2017code,iyer2016summarizing} paradigms. However, the limitation of the token sequence is that it cannot represent the structural features of source code. 

AST (Abstract Syntax Tree), as a tree representation of source code, aims at representing the syntactic structure of source code. 
Many recent methods \cite{leclair2021ensemble,liu2020retrieval,liu2020unified,zhang2019novel, mou2016convolutional, ji2021code} encode ASTs by leveraging GNNs to learn the structural features of source code. 
They obtain the code representation by aggregating nodes in an AST together based on the parent-child connections so that the relationships between the adjacent levels can be captured.
However, the relationships between non-adjacent levels of AST nodes are not given sufficient attention. 
We believe that the relationships between both adjacent and non-adjacent levels are indispensable to represent the overall topology of an AST. 
A major limitation of the existing methods is that they are difficult to fully understand and encode the AST hierarchy since some hierarchical relationships are not considered (we will describe more about the limitation of the existing methods in Section 2).
Therefore, a main challenge is \textit{how to effectively learn the hierarchical topology of ASTs?} This challenge is more acute when the ASTs have deep levels, complex nesting structures, and a large number of nodes. Techniques that can effectively represent such a topology in the embedding space can help neural networks better understand the AST representation of source code.

In this paper, we present HELoC (short for \textbf{H}ierarchical Constrastiv\textbf{E} \textbf{L}earning \textbf{o}f \textbf{C}ode), a hierarchical contrastive learning model for source code representation. To address 
the challenge described above, we propose the following novel techniques:

\textbf{Novelty 1: Contrastive learning for AST hierarchy}. We gain inspiration from the self-supervised contrastive learning (SCL) paradigm \cite{hadsell2006dimensionality}, and discover ways to explicitly preserve the AST hierarchy in the embedding space by formulating hierarchical relationships as distances between node vectors. 
Traditional sequence- and path-based techniques \cite{allamanis2016convolutional,hu2018deep,zugner2020language, alon2019code2vec,alon2018code2seq} can compute the static discrete distance values (e.g., sequence distance or shortest path length) directly from the code and its AST. In contrast, our goal is to design a hierarchical contrastive learning encoder for the AST hierarchy. 
 Therefore, two learning objectives are employed to train HELoC to predict the AST node level and learn the three hierarchical relationships between nodes, which bring the representation vectors of nodes with greater differences in AST levels farther apart in the embedding space.

\textbf{Novelty 2: A specialized GNN for AST hierarchical structures}. 
ASTs are typically characterised by multi-node, multi-level and multi-nested structures. Previous source code representations \cite{fernandes2018structured,mehrotra2021modeling} have demonstrated that GNNs 
can better encode the local structure of ASTs \cite{guo2019attention}. However, 
it is difficult for GNNs to effectively model the long-range dependencies (i.e., global relationships) between nodes of a deep AST. 
In this work, we design a novel GNN called RSGNN (stands for Residual Self-Attention GNN), which utilizes both graph convolutional networks (GCNs) (for capturing local structure) and the self-attention mechanism (for capturing global structure) to learn AST hierarchy comprehensively. 
\begin{figure*}
	\centering
	\includegraphics[width=0.97\linewidth]{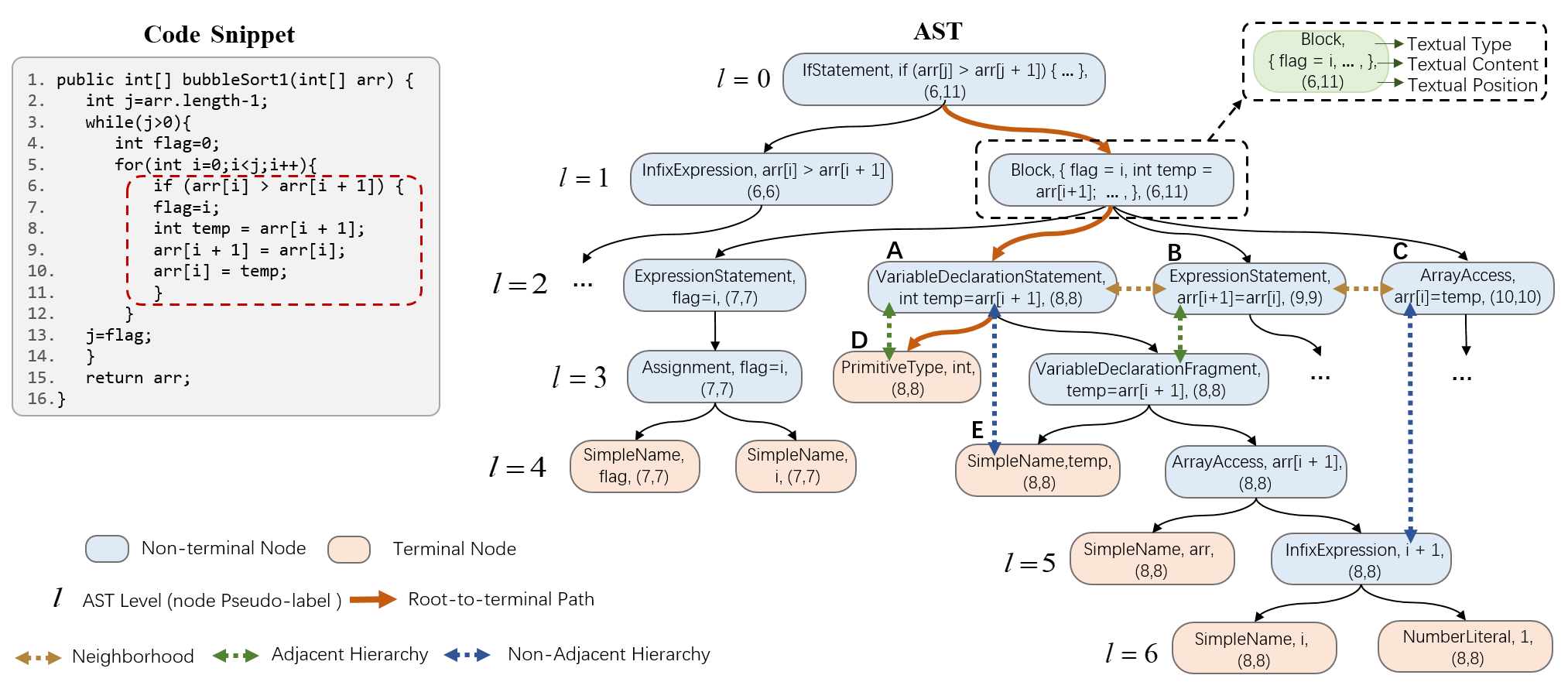}
	\caption{A code snippet and its partial AST.}
	\label{fig:AST}
\vspace{-12pt}
\end{figure*}

We leverage these novel techniques to develop HELoC, a self-supervised, contrastive learning model for source code representation. In HELoC, RSGNN is used as an AST hierarchy encoder and pre-trained under the hierarchical contrastive learning objectives. During pre-training, HELoC learns the node representation and the path representation of the AST and then feeds the combined representation into the RSGNN network. The pre-trained RSGNN can then be used to obtain comprehensive representations of source code in downstream tasks.


We apply our model to three software engineering tasks: code classification, code clone detection, and unsupervised code clustering, and conduct extensive experiments on multiple datasets. The experimental results show that our approach outperforms the respective state-of-the-art models. For code classification, our model improves the accuracy to 97.2\% on GCJ datasets, which is 3.8\% higher than the current state-of-the-art model (ASTNN \cite{zhang2019novel}). For code clone detection, our model improves the F1 scores to 98.0\% and 97.4\% on BCB and GCJ datasets, which is 2.6\% and 4.3\% higher than the current state-of-the-art model (InferCode \cite{bui2021infercode}, FCCA \cite{hua2020fcca}), respectively. For code clustering, HELoC achieves relative improvements of 12.1\% and 12.6\% in ARI metric on OJ and SA datasets when compared to the state-of-the-art method (Infercode \cite{bui2021infercode}).
Our ablation study results also suggest that all components of HELoC are effective. 

Our major contributions are summarized as follows:
\begin{itemize}
\item We propose a novel AST hierarchy representation method based on self-supervised hierarchical constrastive learning, which 
exploits the topological relationships between AST nodes and learn the structures of ASTs more accurately.

\item We present a specialized GNN by constructing internal and external residual self-attention mechanisms, which can capture the AST hierarchy comprehensively. 

\item Our extensive experiments on three downstream tasks confirm the effectiveness of the proposed approach to neural code representation. 


\end{itemize}

\section{Background and Motivation} 







\subsection{AST Hierarchy}
\noindent ASTs are designed to represent the syntactic structure of source code, and it is significant to learn the structural features of ASTs effectively for code representation. An example of AST is shown in Figure ~\ref{fig:AST}. The code snippet implements the bubble sort function. For presentation purpose, we show only a partial AST of the code. 

We consider AST as a planar structure and dissect its structural features in horizontal hierarchy and vertical paths dimensions. Therefore, the AST hierarchy will act as a fundamental feature to study. 
Explicitly, the definition of AST hierarchy includes two aspects: the level of AST nodes and three relationships between node levels.
Level refers to the unique path length from the root to a node. The three types of hierarchical relationships include: 1) \textbf{Neighborhood} -- topological relationships between nodes at the same level; 2) \textbf{Adjacent Hierarchy} -- topological relationships between nodes with a hierarchical difference of 1 and 3) \textbf{Non-Adjacent Hierarchy} -- topological relationships between nodes with hierarchical differences greater than 1. As the AST shown in Figure ~\ref{fig:AST}, the level of nodes $A$, $B$, $C$ is $l=2$, node $D$ is $l=3$, and node $E$ is $l=4$, where the relationship between nodes $A$, $B$, $C$ is neighborhood, nodes $A$, $D$ is adjacent hierarchy, and nodes $A$, $E$ is non-adjacent hierarchy.

\subsection{AST-based Representation Models for Source Code}
\noindent How to represent source code effectively is a fundamental problem in deep learning for software engineering research. 
Recently, many AST-based source code representations have been proposed, which can be roughly divided into two categories: supervised and self-supervised models.

\emph{\textbf{Supervised models.}} Most models of code representation employ a supervised learning paradigm where a specific task (e.g., code classification) is used as the training target, and the code representation is learned based on human-annotated labels. For example, TBCNN \cite{mou2016convolutional} performs supervised learning to obtain code representations by convolving the parent and child nodes through a tree-based convolution kernel, applied to supervised tasks of code classification and similarity detection. 
HAG \cite{ji2021code} aims to distinguish the importance of different nodes by weighted aggregated parent-child nodes in a supervised manner and is used for code clone detection tasks. 
ASTNN~\cite{zhang2019novel} divides each large AST into a sequence of small statement trees and encodes the statement trees by using bottom-up computations from the leaf nodes to the root node to obtain the source code representation, applied to supervised tasks of code classification and code clone detection.

\emph{\textbf{Self-Supervised models.}} Self-supervised learning aims to benefit downstream tasks by designing pretext tasks to learn from large-scale unlabeled data.
The most representative studies include InferCode \cite{bui2021infercode}, Corder \cite{bui2021self}, and Contracode \cite{jain2020contrastive}. InferCode first decomposes the AST of the code into subtrees of unequal granularity, then encodes the nodes of the subtrees with the improved TBCNN. It learns the code representation by predicting subtrees for the pretext task.
Corder enhances the self-supervised framework by recognizing similar and dissimilar code snippets via introducing a contrastive learning mechanism. It is mainly applied to code retrieval and code summarization. Contracode uses source-to-source compiler transformation techniques to generate syntactically different but functionally similar programs for code-level contrastive learning, making the representations of semantically equivalent snippets close to each other, while the representations of non-similar snippets are far apart. 

Progress has also been made in incorporating ASTs into the learning process through self-supervised learning with pre-training models. 
TreeBERT \cite{DBLP:journals/corr/abs-2105-12485} and GraphCodeBERT \cite{guo2020graphcodeBERT} have made good progress as BERT-based models. These models propose novel pre-training tasks to learn the structure of ASTs (e.g., path and data dependencies) to capture more structural information of ASTs. 

\subsection{The Limitations of the Existing Work}
\noindent The two types of source code representation models introduced in previous subsection have four major limitations. \emph{\textbf{1) Heavy reliance on limited labelled data.}} Methods using the supervised learning paradigm rely on human-labelled data. They cannot benefit from the code knowledge embedded in large amounts of unlabelled data, whereas existing self-supervised models effectively alleviate this problem. \emph{\textbf{2) Inadequate learning of AST hierarchy.}} Techniques (such as ASTNN, TBCNN, and HAG) learn the AST structure by aggregating child nodes to their parent node, focusing only on the adjacent hierarchy. Consequently, topologically non-adjacent hierarchy between nodes (sometimes spanning multiple levels) are difficult to learn, resulting in such relationships in the embedding space not being accurately represented. 
Similarly, it is also difficult for existing BERT-based models to learn AST hierarchy.
\emph{\textbf{3) Broken semantics.}} 
As an AST may have many levels and a large number of nodes,
path- and subtree-based techniques, such as Code2vec, Code2seq and Infercode, divide the AST into either a bag of path contexts or sub-trees. However, decomposition of the AST into smaller ones may lose long-range context information, which destroys the overall semantic relationships of the AST and leads to a partial understanding of the overall structure of the code. 
\emph{\textbf{4) Extra training data effort.}}
Corder and Contracode mitigate the problem of broken semantics by applying contrastive learning to syntactically different but semantically similar code snippets. However, this way of obtaining negative code-level samples for contrastive learning is limited by the capacity of the program transformation techniques and the availability of the datasets consisting of similar semantic code snippets. Actually, program transformation techniques cannot exhaust all semantically similar code  
snippets, and it is not easy to obtain sufficient similar semantic code snippets
in practice. Therefore, it is necessary to design a generic model of code representation to enable more comprehensive and autonomous learning.


In this paper, we formulate the learning of AST hierarchy as a pretext task of self-supervised contrastive learning, where cross-entropy and triplet losses are adopted as learning objectives to predict the node level and learn the hierarchical relationships between nodes. 
Being a self-supervised method, our model does not require synthetic negative code samples and decomposition of the ASTs. 

\section{Proposed Model}

\begin{figure}
\centering
\includegraphics[width=1.0\linewidth]{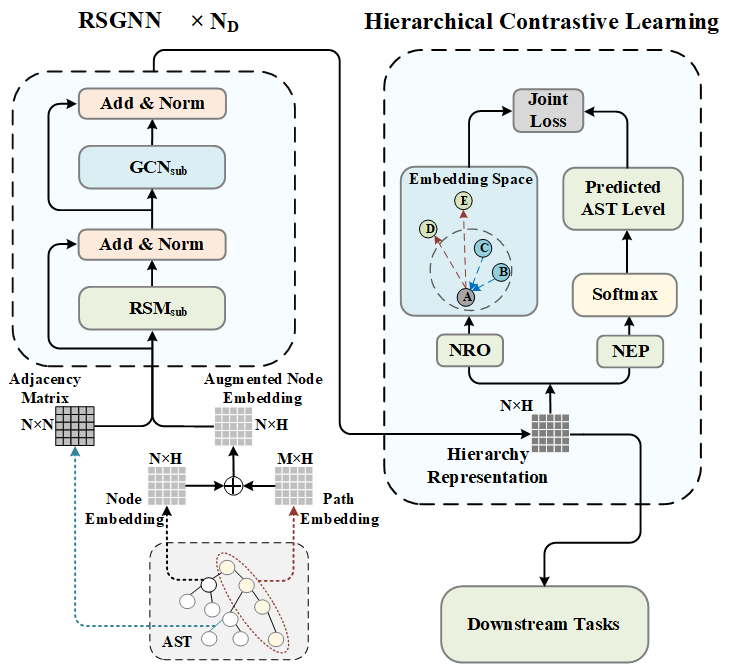}
\caption{An overall framework of the HELoC model. NEP and NRO refer to the two pre-training objective functions of HELoC.} 
\label{fig:HELoC}
\vspace{-6pt}
\end{figure}

\noindent The overall architecture of HELoC is shown in Figure ~\ref{fig:HELoC}. The model accepts the entire AST as input.  RSGNN is then used to encode the AST hierarchy and trained with a joint loss for hierarchical contrastive learning.
The parsed AST (see Figure ~\ref{fig:AST}) contains two types of information: \textbf{1) information represented by the AST nodes,} including the textual type, the textual content, and the textual position (i.e., the startLineNumber and the endLineNumber); and \textbf{2) information represented by the edges between nodes}, i.e., an edge from node $n_i$ to node $n_j$. Note that we utilize the textual content in the nodes because it enables us to extract the syntax of the code while preserving its semantics. In addition, we also record the textual position in each node within code snippet, which helps distinguish between the embedding of the nodes for better contrastive learning.

\subsection{Input Representation}

\noindent We represent an AST with path augmented node embeddings, which incorporate the embedding of the path where each node is located. The details are as follows.


\emph{\textbf{Node Embedding.}} 
We consider an AST node's information (i.e., textual type, textual content and textual position) as a  string, and adopt a Document Embedding technique\footnote{ https://github.com/flairNLP/flair/} to embed the entire string. 
The node embedding is denoted as $X{^{node}_0}=\left\{{x_{n_1},x_{n_2},...,x_{n_N}}\right\} \in R^{N \times H}$ , where $x_{n_i} $ represents the embedding of node $n_i$, $N$ is the number of node, and $H$ is the embedding dimension. Furthermore, we construct an adjacency matrix $A\in R^{N \times N} $ to represent the adjacency relationships  
between nodes, where $A_{ji} = 1$ if node $n_i$ is the parent of $n_j$. 

\emph{\textbf{Path Embedding.}} 
We first extract the paths of an AST from the root node to each terminal node and use the same method (i.e., Document Embedding) used for node embedding to obtain the AST's path initial embedding. An example of path (marked with a thick orange line) is shown in Figure ~\ref{fig:AST}. 
The whole path embedding  $X_{0}^{path}=\left\{x_{p_{1}},x_{p_{2}}, \ldots,x_{p_{M}}\right\}\in R^{M \times H}$ can be obtained,
where $M$ is the number of path, 
and $x_{p_{ij}} $ represents the embedding assigned by $i$-th path to the $j$-th node in the path. 

Finally, the embedding of one path are added to the corresponding node embedding to obtain the augmented node embedding containing the respective path information $X{^{AST}_0}=\left\{{\tilde{x}_{n_1},\tilde{x}_{n_2},...,\tilde{x}_{n_N}}\right\} \in R^{N \times H}$.

\subsection{RSGNN (Residual Self-Attention GNN)} 
\noindent GNN-based representations tend to be local and struggle to leverage long-range interactions. On the contrary, self-attention allows global information flow at each step. Therefore, we combine the advantages of both in the design of RSGNN and add residual connection to alleviate gradient vanishing due to encountering many iterations on deep AST.

RSGNN consists of $N_D$ layers, where each layer contains two different sub-layers, i.e., a residual self-attention mechanism sub-layer ($RSM_{sub}$) and a graph convolution sub-layer ($GCN_{sub}$). 
We add residual connection and layer normalization between the sub-layers. 

\begin{figure}
	\centering
	\includegraphics[width=1.0\linewidth]{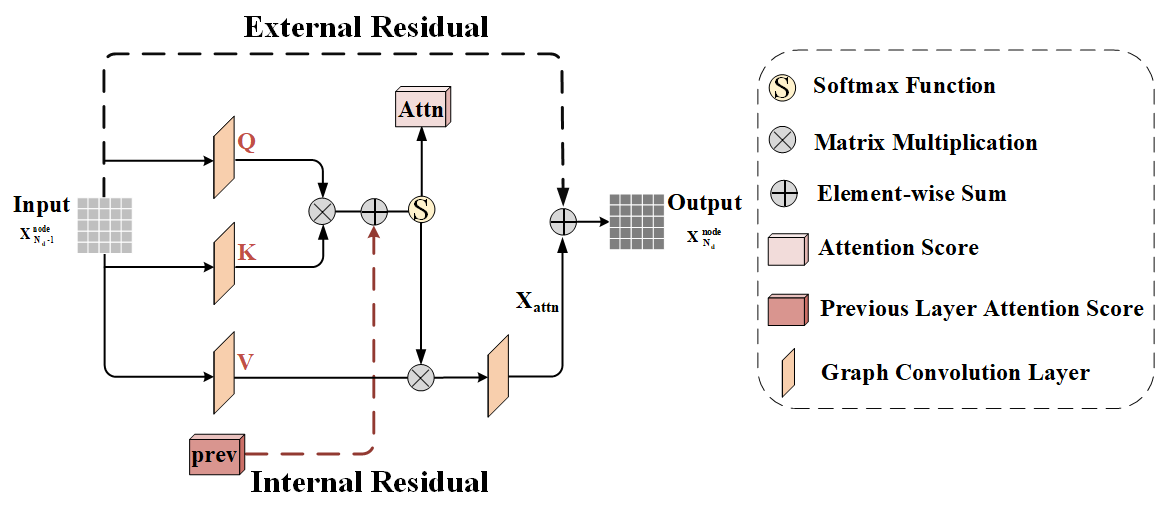}
	\caption{Residual Self-attention Mechanism.}
	\label{fig:RSM}
\vspace{-12pt}
\end{figure}

\emph{\textbf{Residual self-attention mechanism sub-layer.}} The $RSM_{sub}$ is described in Figure \ref{fig:RSM}. First, we take $X{^{AST}_{N_{d}-1}}$ (the previous RSGNN’s output) as input and perform graph convolution mappings on it to compute the \textit{query, key}, and \textit{value} , i.e., \textit{Q, K,} and \textit{V}, respectively. The first layer's input is $X{^{AST}_0}$ and we formulate this process as follows:
\begin{equation}
Q=GCN\left(X{^{AST}_{N_{d}-1}}, \tilde{A}\right)=\sigma\left(D^{-1} \tilde{A} X{^{AST}_{N_{d}-1}} W_{Q}\right)
\end{equation}
where $\tilde{D}_{i i}=\sum_{j} \tilde{\mathbf{A}}_{i j}$ represents the diagonal matrix, and $W_Q$ is the learnable weight matrix. $K$ and $V$ are calculated as the same way.



During the network training, gradient vanishing sometimes occurs due to too many stacked layers, and we solve this problem by adding internal residual connection and external residual connection to self-attention mechanism. 

We form the \textbf{\emph{internal residual connection}} by adding the attention score of the previous layer as one additional input to the attention score calculation of the current layer.
\begin{equation}
Attn=softmax\left(\frac{QK^\top}{\sqrt{d}}+prev\right)  
\end{equation}
where $prev$ represents the attention score of the previous layer, and $Attn$ will be passed to the next layer as its previous layer's attention score.
Furthermore, we feed $Attn$ and $V$ to GCN and get the output $X_{attn}$.
\begin{equation}
\begin{split}
X_{attn} &=GCN((Attn,V),\tilde{A})
\\&= \sigma\left(D^{-1} V\tilde{A} \left(softmax\left(\frac{QK^\top}{\sqrt{d}} +prev\right)\right) \cdot W_l \right)
\end{split}
\end{equation}
where $W_l$ is the learnable weight matrix.

Finally, the \textbf{\emph{external residual connection}} is introduced to obtain the final output of $RSM_{sub}$. Specifically, we establish a residual connection between the output $X_{attn}$ of self-attention and the previous $RSM_{sub}$ layer’s output.
\begin{equation}
{RSM_{sub}}\left(X_{N_{d}-1}^{\text {AST }}\right)=X{^{AST}_{N_{d}-1}}+X_{attn}
\end{equation}

\emph{\textbf{Graph convolution sub-layer.}} $GCN_{sub}$ consists of two GCNs with a RELU activation in between. 

The output of the $N_d$-th RSGNN layer is calculated as follows:
\begin{equation}
\begin{gathered}
\widetilde{X_{N_d}^{\text {AST }}}=\text { LayerNorm }\left(X_{N_{d}-1}^{\text {AST }}+{RSM_{sub}}\left(X_{N_{d}-1}^{\text {AST }}\right)\right) \\
X{^{AST}_{N_{d}}}=\text {LayerNorm}\left(\widetilde{X{^{AST}_{N_{d}}}}+GCN_{sub}\left(\widetilde{X_{N_{d}}^{\text {AST} }}, \tilde{A}\right)\right)
\end{gathered}
\end{equation}
where $X{^{AST}_{N_{d}-1}}$ is the previous RSGNN’s output, $\tilde{\mathbf{A}}=\mathbf{A}+\mathbf{I}_{N}$ represents the adjacency matrix with added self-connections, $\mathbf{I}_{N}$ is the identity matrix.

After $N_D$ layers of RSGNN, we obtain the representation of the AST nodes, which is denoted as $X_{N_D}^{AST}=\left\{\bar{x}_{n_1},\bar{x}_{n_2},...,\bar{x}_{n_N}\right\} $ . 
\subsection{Hierarchical Contrastive Learning} 
\noindent 
In this work, we propose a novel hierarchical contrastive learning (HCL) method, which takes the learning of AST hierarchy as the pretext task \cite{misra2020self} for contrastive learning. 
We use the construction of triplets in Figure~\ref{fig:AST} to 
explain the idea of our HCL.
Any AST nodes can be considered an anchor node, e.g. node $A$. Nodes at the same level as the anchor node are treated as the positive nodes, e.g., nodes $B$ and $C$, while nodes at other levels are treated as the negative nodes (including those at the adjacent levels, e.g., node $D$, and those at the non-adjacent levels, e.g., node $E$).
On this basis, we can construct the triplets (e.g., $\{A, B, D\}$, $\{A, C, E\}$), as well as positive pairs (e.g., $<A, B>$, $<A, C>$) and negative pairs (e.g., $<A, D>$, $<A, E>$). Contrastive learning encourages the representation of the anchor node to be closer to the representations of the ``positive” nodes, while staying away from the representations of ``negative” nodes. For an AST hierarchy, the representations of nodes with greater differences in AST levels should be farther apart in the embedding space. For example, $<A, E>$ should be more distant than $<A, D>$ as the nodes $A$ and $E$ have a larger difference in levels. However, the original contrastive learning objective is insufficient for our HCL to learn the AST hierarchy.


To address this issue, we take the AST levels as pseudo-labels and adopt two objective functions: 1) node level prediction (NEP) to predict the AST levels where the nodes are located, and 2) node relationship optimization (NRO) to learn the three topological relationships between the nodes. In particular, we need to consider the following two aspects in NRO: one is to construct ``positive" and ``negative" pairs according to the node hierarchy indicated by the pseudo-labels; the other is to 
inject into the learning objective the difference value between the level of the negative node and that of the anchor node. In this way, the distance between node embeddings is continuously enlarged as their hierarchical differences increase, thus achieving the goal of learning the intrinsic topological features that distinguish the current AST from others. 
We elaborate the above learning process as follows.

\emph{\textbf{AST pseudo-label construction.}} To learn the hierarchical structure, we use the given AST’s level to construct pseudo-labels for the nodes automatically without human annotations so that our approach can be self-supervised. Specifically, we use the depth first search (DFS) algorithm to traverse the AST nodes, labelling the root node as 0, the first-level nodes as 1, the second-level nodes as 2, and so on, with the $l$-th-level nodes labeled as $l$ (see Figure ~\ref{fig:AST}). 

\emph{\textbf{Hierarchy representation learning.}} We pre-train HELoC by predicting the AST hierarchy through contrastive learning. For this purpose, we implement a single-layer linear neural network as a classifier to predict the level of AST node and obtain: $\hat{y}=\left(W_{AST}X_{N_D}^{AST}+b_{AST}\right)$, where $W_{AST}$ is the weight matrix, $b_{AST}$ is the bias term, $X_{N_D}^{AST}$ refers to the node representation of the RSGNN output.

The pre-training objectives of HELoC include NEP and NRO. NEP employs a cross-entropy loss function~\cite{rubinstein1999cross}, denoted as $\mathcal{L}_{h}$, and NRO employs a triplet loss function \cite{schroff2015facenet}, denoted as $\mathcal{L}_{t}$. We use a joint loss function~\cite{wang2021code} for HELoC to automatically balance $\mathcal{L}_{h}$ and $\mathcal{L}_{t}$ during the learning process without human intervention, which is defined as:
\begin{equation}
\begin{split}
\mathcal{L} &\approx\frac{1}{\theta^{2}} \mathcal{L}_{h}+\frac{1}{\tau^{2}} \mathcal{L}_{t}+\log \theta+\log \tau
\\& =\exp \left(-2 \theta^{\prime}\right) \cdot \mathcal{L}_{h}+\exp \left(-2 \tau^{\prime}\right) \cdot \mathcal{L}_{t}+\theta^{\prime}+\tau^{\prime}
\end{split}
\end{equation}
\begin{equation}
\mathcal{L}_{h}=\sum_{i=1}^{N}\left(-\log\frac{\exp \left(y{_{i}^{level}}\right)}{\sum_j \exp \left(y{_{j}^{level}}\right)}\right)
\end{equation}
\begin{equation}
\mathcal{L}_{t}=\sum_{i=1}^{N}\left[\left\|\bar{x}_{n_{i}}^{anc}-\bar{x}_{n_{i}}^{pos}\right\|_{2}^{2}-\left\|\bar{x}_{n_{i}}^{anc}-\bar{x}_{n_{i}}^{neg}\right\|_{2}^{2}+	\Delta{l}+\alpha\right]_+
\end{equation}
where $\theta$ and $\tau$ are learnable scalars and $N$ is the number of nodes in an AST. $y{_{i}^{level}}$ is the probability that the predicted $i$-th node level is $l$. $\left \langle anc,pos,neg\right \rangle$ denotes a triplet, in which $anc$ is the anchor node, $pos$ is the positive node, $neg$ is the negative node, and $\bar{x}_{n_{i}}^{(\cdot)}$ refers to the representation of the nodes of the constructed triplet. $\Delta{l}$ represents the difference in levels between the negative node and the anchor node, aiming to make the representation of nodes with more distant levels in an AST more distant in the embedding space. 
$\left\|*\right\|_{2}^{2}$ is the Euclidean distance metric, where $\alpha$ is a margin that is enforced between positive and negative pairs. $[]_+$ indicates that if the value within $\left[\right]$ is greater than zero, the value is treated as a loss; if it is less than zero, the loss is zero. Note that, for numerical stability, we let $\theta^{\prime}=\log \theta, \tau^{\prime}=\log \tau$ and train HELoC to learn  $\theta^{\prime}$ and $\tau^{\prime}$ instead of the unconstrained $\theta$ and $\tau$ \cite{kendall2018multi}. 

\begin{figure}[htbp]
	\centering
	\includegraphics[width=1.0\linewidth]{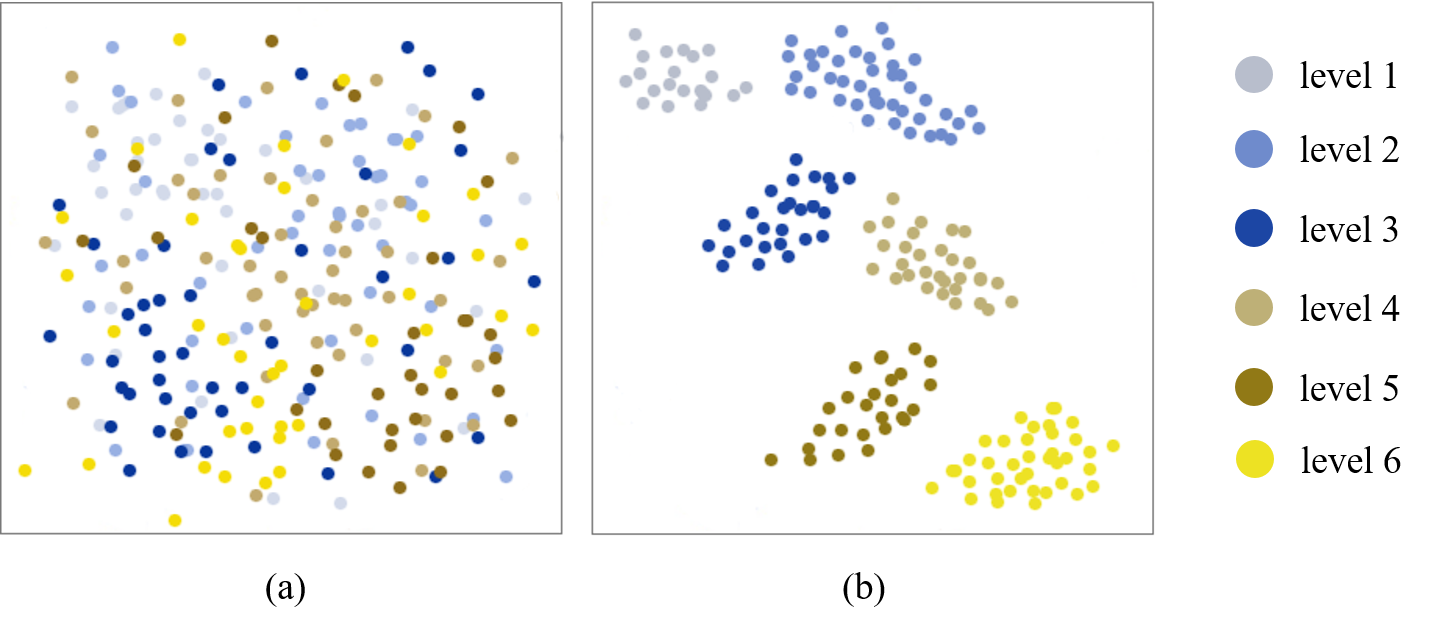}
	\caption{Visualization of the hidden vector of a six-level AST. 
	The points in the figure corresponds to the AST nodes.}
	\label{t-SNE}
\end{figure}

To visualize the results of HELoC, we plot the t-distributed stochastic neighbor embedding (t-SNE) \cite{van2008visualizing} diagrams of the hierarchy representation, as shown in Figure ~\ref{t-SNE}. We select the middle six levels of an AST {(from our pre-training Java-Large dataset described in Section 5.1)} as an example.
Figure ~\ref{t-SNE} (a-b) shows the output before and after using HELoC (both using K-means clustering), respectively. HELoC leads to the clustering of nodes in the same level, while nodes with greater differences in AST levels farther apart in the embedding space.
This validates the ability of HELoC to learn AST hierarchy information.

\section{Applications of the Proposed Model}

\noindent HELoC can be applied in two ways: 1) fine-tuning, where HELoC acts as a pre-trained model and its pre-trained parameters can be fine-tuned in supervised learning tasks, and 2) producing code vectors, where HELoC serves as a feature extractor without fine-tuning any parameters, and the resulting code representations are directly used as input to a model that accomplishes a specific downstream task.

\emph{\textbf{Fine-tuning for Code Classification.}} Code classification refers to classifying source code according to their functionalities. Given the representation $r$ of a code snippet and its functionality label $cla$, we can classify it via a fully-connected layer (with a total of $c$ classes). We define the loss function as the widely used cross-entropy function:

\begin{equation}
Loss_{cla}=-\log \frac{\exp \left(y{_{i}^{cla}}\right)}{\sum_{j} \exp \left(y{_{j}^{cla}}\right)}
\end{equation}
where $y^{cla}=W_{0} r+b_{0}$, and $W_0$ is the weight matrix, $b_0$ is the bias term.

For the prediction phase, the output $y^{cla}$ indicates the predicted probability that code snippet belongs to the corresponding class.
\begin{equation}
 prediction = argmax\left(y^{cla}\right), i=1, \ldots, c
\end{equation}

\emph{\textbf{Producing Code Vectors for Code Clone Detection.}} Code clone detection, i.e., detecting whether two code snippets are similar or not, is a widely studied topic in software engineering research \cite{sajnani2016sourcerercc,white2016deep,tufano2018deep}. Given two code snippets $c_1$ and $c_2$, the code representations $r_1$ and $r_2$ are produced by HELoC. We use a classifier (a linear layer) for the secondary training, the mean squared error (MSE) is used as the loss function:
\begin{equation}
Loss_{clo}=\left(y{_{i}^{clo}}-\frac{r_{1} \cdot r_{2}}{\left|r_{1}\right| \cdot\left|r_{2}\right|}\right)^{2}
\end{equation}
where $\frac{r_{1} \cdot r_{2}}{\left|r_{1}\right| \cdot\left|r_{2}\right|}$ is used to denote the relatedness \cite{tai2015improved} between $r_1$ and $r_2$. $y{_{i}^{clo}}$ is set to 1 and -1 for the true clone pairs and false clone pairs, respectively.

For the prediction phase, the output $p$ (i.e., $\frac{r_{1} \cdot r_{2}}{\left|r_{1}\right| \cdot\left|r_{2}\right|}$) is in the range [-1,1]. Thus, the prediction can be obtained as follows:
\begin{equation}
prediction =\left\{\begin{array}{ll}
 True,  & p>0\\
 False,  & p \leq 0
\end{array}\right.
\end{equation}

{\emph{\textbf{Producing Code Vectors for Code Clustering.}} Code clustering is an unsupervised task of automatically clustering similar code fragments into the same groups~\cite{bui2021infercode}. Given the vector representation $r$ of a code snippet, we can classify it by a Euclidean distance based similarity measure and the K-means~\cite{kanungo2002efficient} clustering algorithm.}

\section{Experiments}
\noindent This section first presents the pre-training 
of HELoC and then 
the applications HELoC to three program comprehension tasks (code classification, code clone detection, and code clustering). We compare HELoC with state-of-the-art methods. Furthermore, we validate the usefulness of the core components of the model through ablation experiments. 

\subsection{Model Pre-training}
\noindent We pre-train HELoC on the Java-Large dataset \cite{alon2019code2vec}, which contains about 4 million Java source files collected from GitHub. 
HELoC is implemented with PyTorch$ \footnote{ https://pytorch.org/} $ and Deep Graph Library (DGL)$ \footnote{ https://www.dgl.ai/} $. 
In the experiments, the initial input dimension is set to 768. The number of RSGNN layers is set to $N_D=4$. 
Considering the GPU memory and training time, we set the hidden state length of the cache to 256.
The max AST depth, the max path number per AST, and the max node number are set to 30, 200, and 1000, respectively, based on the statistical information about the datasets.
The Adam optimizer \cite{kingma2014adam} is used to train our model with a learning rate of 1e-4 and the batch size is 2048. 

\subsection{Code Classification}
\subsubsection{Datasets and Settings.} We evaluate HELoC on two datasets, i.e., Google Code Jam (GCJ) \cite{liang2018automatic} and Online Judge (OJ) \cite{mou2016convolutional}. The statistics of the two datasets are listed in Table 1. 
\begin{table}[!htbp]
	\centering
	\caption{ Statistics of GCJ and OJ datasets.}
	\vspace{-6pt}
	\label{t1}
	\setlength{\tabcolsep}{6.0mm}{
		\begin{tabular}{c c c }
		 	\toprule \textbf{ Dataset }    &\textbf{GCJ}  &\textbf{OJ}   \\ 
			\hline
	  	\textbf{$\sharp$Code snippets}      & 8647  & 52,000       \\
		\textbf{$\sharp$Classes}       & 6      & 104           \\
		\textbf{Avg. AST nodes}        & 387.3  & 190          \\
		\textbf{Max AST nodes}         & 1,059  & 7,027        \\
		\textbf{Avg. AST depth}        & 33.9   & 13.3         \\
		\textbf{Max AST depth}         & 208    & 76        \\
			\toprule
		\end{tabular}  
	}
\end{table}

\textbf{GCJ.} This dataset is collected from an online program competition held annually by Google. We choose 8647 Java class files from the dataset published by Liang et al. \cite{liang2018automatic} and classify them into six classes according to the problems they solve.

\textbf{OJ.} This dataset consists of C programs collected from the OJ programming exercise system. It is publicly available from Mou et al.$ \footnote{  https://sites.google.com/site/treebasedcnn/} $, which contains 104 classes of problems.
The dataset consists of the programming task and the corresponding source code submitted by the students. If the programs aim to solve the same problem, they have the same functionality and category.

In the experiments, the number of epochs is set to 100 and early stopping is adopted. 
Learning rate and batch size are set to 1e-5 and 64, respectively. The ratio of training, validation, and test sets is set to 8:1:1.
In addition, label smoothing \cite{muller2019does} is adopted to retain values with labels of 0, with a probability $\lambda$, where $\lambda$ is empirically set to 0.1.
We use javalang$ \footnote{  https://github.com/c2nes/javalang} $ and pycparser$ \footnote{https://pypi.python.org/pypi/pycparser} $ to parse the source code written in Java and C and extract the ASTs.

\subsubsection{Baselines and Evaluation Metrics.} We consider three types of baselines that parse code from different perspectives, namely, token-based models, tree-based models, and graph-based models. 
The implementation of the baselines and their parameter settings are provided by their original papers. 

\emph{\textit{Token-Based Model: }}
Previous work \cite{zaremba2014learning,cummins2017synthesizing,iyer2016summarizing}uses \textbf{LSTM} (long short-term memory) networks to model the input sequence of code tokens.

\emph{\textit{Tree-Based Models: }}
\textbf{Tree-LSTM} \cite{tai2015improved} is a generalization of LSTM for modelling tree structure.
\textbf{CodeRNN} \cite{liang2018automatic} designs a new AST-based recurrent neural network by going through the tree from leaf nodes to root node to obtain the final code representation vector.
\textbf{TBCNN} \cite{mou2016convolutional} is a tree-based convolutional neural network model that designs a tree-based convolution kernel consisting of a set of subtree feature detectors that slide the entire AST to extract the structural information of the code.
\textbf{ASTNN} \cite{zhang2019novel} is an AST-based neural network that divides a complete AST into a sequence of small statement trees to solve the long-term dependency problem caused by large ASTs.
\textbf{InferCode} \cite{bui2021infercode} applies self-supervised learning to the AST by predicting subtrees automatically identified from the AST context for code representation.

\emph{\textit{Graph-Based Models: }}
\textbf{GGNN} \cite{allamanis2017learning} uses graphs that attach additional dataflow edges to ASTs to represent the structure of the source code, and then learns code representation by applying gated GNN (GGNN).

To evaluate the effectiveness of code classification models, we use the Accuracy metric, which calculates the percentage of the test set that is correctly classified.

\subsection{Code Clone Detection}

\noindent There are generally four different types of code clones \cite{sajnani2016sourcerercc,white2016deep,roy2007survey}, including Type-1 clones (a code pair that is syntactically identical except for spaces and comments), Type-2 clones (a code pair that is identical on a Type-1 basis except for variable names, type names, and function names), Type-3 clones (a code pair that has several statement additions and deletions but remains syntactically similar, except for the cloning differences between Type-1, Type-2), and Type-4 clones (a code pair with the same functionality that is not textually or syntactically similar but has semantic similarities).

\subsubsection{Datasets and Settings.} We evaluate HELoC on three datasets: BCB, GCJ, and OJClone. Table 2 provides the statistics.
\begin{table}[!htbp]
	\centering
	\caption{ Statistics of the datasets for code clone detection.}
	\vspace{-6pt}
	\label{t3}
	\setlength{\tabcolsep}{3.8mm}{
		\begin{tabular}{c c c c }
			\hline  \textbf{ Dataset }  &\textbf{BCB}  &\textbf{GCJ}  &\textbf{OJClone}   \\ 
		    \hline 	\textbf{$\sharp$Clone pairs}  &97,535  &50,000  &50,000    \\ 
			\textbf{ \%True clone pairs} &95.7\%  &20.0\%  &6.6\%  \\
		    \textbf{ Avg. code length} &27.1  &55.3  &35.4  	\\
	    	\textbf{ Avg. AST nodes} &206  &295.0  &192  	\\	
			\textbf{ Max AST nodes} &15,217  &991  &1,624  	\\
			\textbf{ Avg. AST depth} &9.9  &15.2  &13.2  	\\
			\textbf{ Max AST depth} &192  &27  &60  	\\
			\toprule
			
		\end{tabular}  
	}
\end{table}

\textbf{BCB.} In our experiments, we use the BCB dataset published by Zhang et al. \cite{zhang2019novel}, which is a subset of the BCB constructed by Svajlenko et al.~\cite{svajlenko2014towards}. 
BCB is suitable for evaluating the semantic clone detection capabilities of models. 
Note that in this dataset, Type-3 clones are further divided into strong Type-3 (ST3) and moderately Type-3 (MT3). 

\textbf{GCJ.} The GCJ clone dataset \cite{liang2018automatic} consists of 50,000 clone pairs from 12 different competing problems. Since few code snippets are syntactically similar in these competing problems, most code pairs from the same problem are Type-4 clones.

\textbf{OJClone.} We use the OJClone dataset provided in \cite{zhang2019novel}. Each code snippet in the dataset is a C function. The dataset consists of 50,000 randomly selected samples, produced from 500 programs that were selected from each of the first 15 programming problems in OJ \cite{mou2016convolutional}.
Two code snippets dealing with the same problem form a clone pair, while code snippets solving different problems are not clones. 

The parameter settings are consistent with the code classification (see Section 5.2.1).
In order to make the comparison as fair as possible, for all the baselines, we adopt the implementation and parameter settings provided by their original papers.

\subsubsection{Baselines and Evaluation Metrics.}  To systematically evaluate the proposed model, we compare it with token-based models, tree-based models, and graph-based clone detection models. 

\emph{\textit{Token-Based Models: }}
\textbf{RtvNN} \cite{white2016deep} uses an RNN to learn the representations of code tokens and trains a recursive autoencoder to learn AST representations.
\textbf{SourcererCC} \cite{sajnani2016sourcerercc} is a well-known lexical-based clone detector.

\emph{\textit{Tree-Based Models: }}
\textbf{Deckard} \cite{jiang2007deckard} is a classical AST-based clone detector by identifying similar subtrees to learn tree representations of source code. 
\textbf{CDLH} \cite{wei2017supervised} learns code representations via AST-based LSTM and uses a hash function to optimize the function similarity between code snippets.
Furthermore, \textbf{TBCNN} \cite{mou2016convolutional}, \textbf{ASTNN} \cite{zhang2019novel}, and \textbf{InferCode} \cite{bui2021infercode} (see Section 5.2.2) are also compared on this task.

\emph{\textit{Graph-Based Models: }}
\textbf{SCDetector} \cite{wu2020scdetector} combines the advantages of the token-based approach and the graph-based approach to detect clones with similar software functionality.
\textbf{FA-AST} \cite{wang2020detecting} constructs a flow-augmented AST by adding explicit control and data flow edges, thus enabling the method to make full use of the semantic information for code representations.
\textbf{FCCA} \cite{hua2020fcca} is based on a hybrid representation that fuses heterogeneous structured and unstructured code information from text, AST, and control flow graph (CFG) representations. 

Since code clone detection can be formulated as a binary classification problem (clone or not), we choose the commonly used Precision (P), Recall (R), and F1 scores (F1) as the evaluation metrics.

{\subsection{Code Clustering }

\noindent Code clustering is to group source code according to their similarity. 
Unlike code classification (a supervised task), code clustering is an unsupervised method. It is used as a downstream task by the related work InferCode~\cite{bui2021infercode}. In our work, we also evaluate the effectiveness of HELoC in code clustering. 

\subsubsection{Datasets and Settings.} We evaluate HELoC on two datasets, i.e., Online Judge (OJ)~\cite{mou2016convolutional} and Sorting Algorithm (SA)~\cite{bui2019bilateral}. 

\textbf{OJ:} 
The dataset is the same as the one used in the code classification task. We use the K-means algorithm (K=104) to cluster the code into 104 clusters, following the settings described in InferCode~\cite{bui2021infercode}. 

\textbf{SA:} This dataset 
includes a 10-class sorting algorithm written in Java. We use the K-means algorithm (K=10) to cluster the code into 10 clusters, following the settings described in InferCode~\cite{bui2021infercode}. 

\subsubsection{Baselines and Evaluation Metrics.}

For code clustering, our method does not require labels for training, and is used in an unsupervised manner. However, the previous methods such as ASTNN and TBCNN were trained and validated on the classification task with labels, which violates the prerequisite that the clustering task is label-unaware. Therefore, for the sake of fairness, in this task, we compare our approach with the following typical related methods that support unsupervised tasks. The related methods are used as encoders to generate code vector representations, over which code clustering is performed.

\emph{\textit{Token-Based Model: }}
\textbf{Word2vec}~\cite{mikolov2013distributed} and \textbf{Doc2vec}~\cite{le2014distributed} are well-known baseline methods in NLP, which extract feature vectors by parsing the source code into text.

\emph{\textit{Tree-Based Model: }}
\textbf{Code2vec}~\cite{alon2019code2vec} parses a code fragment into a collection of AST paths. The core idea is to use a soft-attention mechanism on the paths and aggregate all vector representations into a single vector to predict the method name. 
\textbf{Code2seq}~\cite{alon2018code2seq} is similar with Code2vec, which represents a code fragment as a set of combined paths in an AST and uses attention to select the relevant paths when decoding. The vector representations is used to generate the code textual summary.
\textbf{InferCode}~\cite{bui2021infercode} (see Section 5.2.2) is also compared on this task.

\emph{\textit{Graph-Based Model: }}
\textbf{GGNN}~\cite{allamanis2017learning} (see Section 5.2.2) is also compared in this task.

To evaluate the code clustering effectiveness of HELoC, we use Adjusted Rand Index (ARI)~\cite{santos2009use} as the metric. Specifically, we cluster all code snippets based on the similarity between code vectors without class labels. Then, we evaluate the effect of clustering by calculating the distribution similarity function of the class labels in the dataset and the class labels obtained by clustering. 
}

\subsection{Research Questions and Results}

\noindent\textbf{RQ1: How does our approach perform in code classification?}
The results are shown in Table 3. Overall, HELoC improves the accuracy from 93.4\% to 97.2\% on GCJ, and from 98.2\% to 99.6\% on OJ, in comparison with the best competitor (ASTNN). In particular, the greatest improvement was seen on the GCJ dataset with more AST nodes and greater depth, which strongly confirms that HELoC is better to capture this hierarchical complexity of the AST.
We also analyse the performance of the compared methods as follows.

\textbf{Token-based approach}, LSTM performs poorly in our experiments because it uses only the literal meanings of the token words to distinguish code functionalities. 
Since some token words used in OJ are mostly arbitrary, e.g., the names of identifiers are usually $i$, $j$, $a$, etc, most tokens contribute little to the identification of code features. 
The experimental results show that the token-based approach performs the worst compared to other types of approaches. This is because it aims at modelling the token sequence, which leads to a loss of syntactic structure information of the code.

\textbf{Tree and graph-based approaches}, such as CodeRNN, TBCNN, and ASTNN learn the representation of code in various ways. In particular, TBCNN performs better than CodeRNN because it can encode the features of the given tree structure more accurately by designing tree-based convolution kernel for the AST. 
ASTNN does not perform as well as HELoC. 
One reason is that ASTNN uses bottom-up aggregation of leaf nodes to the root node, which does not capture the non-adjacent hierarchy.
Another reason could be that ASTNN decomposes an AST into a series of statement subtrees in depth-first traverse order so that different relationships such as nesting and if/else branches between statements may be lost. HELoC can handle the case of a huge AST without splitting the AST, avoiding this problem.
In addition, an unexpected finding is that GGNN performs poorly on the code classification, with even lower accuracy than the token-based model. This may be because GGNN, as a graph-based model, views the AST as an ordinary graph and focuses more on the data flow between the nodes, ignoring the unique hierarchical structure of AST.
\begin{table}[tbp]
	\centering
	\caption{ Code classification accuracy (\%) on GCJ and OJ.}
		\vspace{-6pt}
	\label{t4}
	\setlength{\tabcolsep}{3.8mm}{
		\begin{tabular}{c c c c }
		 \toprule 
			 \textbf{ Groups }  &\textbf{ Methods }    &\textbf{GCJ}  &\textbf{OJ}   \\ 
			\hline
			\multirow{1}{*}{\textbf{Token-based}} 
			&LSTM       & 78.4    & 88.0 \\
				\hline
			 	\multirow{5}{*}{\textbf{Tree-based}} &Tree-LSTM  & 88.9    & 88.2        \\
			 &CodeRNN    & 87.4    & 90.2         \\
			 &TBCNN       & 91.7    & 94.0        \\
			 &ASTNN       & \textbf{93.4}    &\textbf{ 98.2 }       \\
			 &InferCode  & 90.8    & 96.3        \\
			 	\hline	
			 		\multirow{1}{*}{\textbf{Graph-based }} &GGNN       & 72.9    & 79.5        \\
		\hline
		\multirow{1}{*}{\textbf{Our approach}} &\textbf{HELoC} &\textbf{ 97.2}   &\textbf{ 99.6}        \\	
			\toprule
		\end{tabular}  
	}
	\vspace{-6pt}
\end{table}

\noindent\textbf{RQ2: How does our approach perform in code clone detection?  }
Tables 4-5 show the precision, recall, and F1 obtained by our approach on BCB, GCJ, and OJClone datasets. Our approach scored better than all baselines, and the F1 scores are improved from 95.4\% (InferCode) to 98.0\% on BCB, from 93.1\% (FCCA) to 97.4\% on GCJ and from 95.7\% (ASTNN) to 98.0\% on OJClone. In the following, we analyse the performance of compared methods in detail. 
\begin{table*}[!htbp]
	\centering
	\caption{Results of code clone detection on BCB.}
		\vspace{-6pt}
	\label{t5}
	\setlength{\tabcolsep}{1.4mm}{
		\begin{tabular}{c c c c c c c c c c c c c c c c c c c c c c c c}
			\toprule
		\multirow{2}{*}{\textbf{Type}}   & \multicolumn{3}{c}{\textbf{ASTNN}} &&	\multicolumn{3}{c}{\textbf{FA-AST }} &	&\multicolumn{3}{c}{\textbf{SCDetector }} &	&\multicolumn{3}{c}{\textbf{FCCA }} & &	\multicolumn{3}{c}{\textbf{InferCode }}&	 &\multicolumn{3}{c}{\textbf{Our approach}}\\
		\cmidrule(r){2-5} \cmidrule(r){6-9} \cmidrule(r){10-13} \cmidrule(r){14-17} \cmidrule(r){18-21} \cmidrule(r){22-24}
	& P	& R   & F1   && P	& R   & F1  && P 	& R    & F1 & & P  	& R     & F1  & & P & R  & F1 & & P	& R  & F1         \\ 
			\hline
	     BCB-T1    & 100    & 100   & 100 && 100 & 100 & 100 && 100 & 100 & 100 && 100 & 100 & 100 && 100 & 100 & 100 && 100 & 100 & 100        \\
		 BCB-T2   & 100    & 100  & 100 && 100 & 100 & 100 && 100 & 100 & 100 && 100 & 100 & 100 && 100 & 100 & 100 && 100 & 100 & 100        \\
		 BCB-ST3   & 99.9   & 94.2 & 97.0 && 100 & 99.6 & 99.8 && 100 & 94.6 & 97.2 && 100 & 99.8 & 99.9  && 99.8 & 94.2 & 97.0 && 100 & 100 & 100        \\
		BCB-MT3    & 99.5   & 91.7 & 95.5 && 98.7 & 96.5 & 97.6 && 97.7 & 95.9 & 96.8 && 98.7 & 95.9 & 97.3   && 99.5  & 96.2  & 97.8  && 99.9 & 98.0 & 98.9        \\
		 BCB-T4     & 99.8   & 88.3 & 93.7 && 97.7 & 90.5 & 94.0 && 97.3 & 92.3 & 94.7 && 98.2 & 92.4 & 95.2   &&98.2  & 92.7  &95.3  && 99.8 & 96.2 & 98.0        \\
		\hline
		 BCB-ALL   & \textbf{99.8}  & 88.4 & 93.8 && 97.7 & 90.6 & 94.0 && 97.3 & 92.4 & 94.8 && 98.2 & 92.4 & 95.2   && 98.2  & \textbf{92.8}   & \textbf{95.4}  & & \textbf{99.8} & \textbf{96.2} & \textbf{98.0}       \\
		\toprule
		\end{tabular}
	}
\end{table*}

\begin{table}[!htbp]
	\centering
	\caption{Results of code clone detection on GCJ and OJClone.}
		\vspace{-6pt}
	\label{t6}
	\setlength{\tabcolsep}{0.5mm}{
		\begin{tabular}{ c c c c c c c c c }  
				\toprule
			\multirow{2}{*}{\textbf{Groups}}  &\multirow{2}{*}{\textbf{Methods}}	&\multicolumn{3}{c}{\textbf{GCJ }}   && \multicolumn{3}{c}{\textbf{OJClone}} \\	
			\cmidrule(r){3-5} \cmidrule(r){7-9}	 & & P  & R     & F1   	&& P   	& R     & F1        \\ 
			\midrule
			\multirow{2}{*}{\textbf{Token-based}} 
			&RtvNN   &20.4 &90.0 &33.3 &&36.1 &83.3 &50.4  \\
			&SourcererCC   &43.4 &11.2 &17.8 &&7.4  &74.2 &13.5  \\
			\hline
			\multirow{5}{*}{\textbf{Tree-based}} 	&Deckard   &45.0  &44.2 &40.1 &&\textbf{99.0} &5.3 &10.1  \\
			&TBCNN   &91.7  &89.3 &90.5 &&90.1 &81.3 &85.8  \\
			&CDLH   &46.1 &69.8 &55.5 &&47.2 &73.3 &57.4  \\
			&ASTNN   &95.4 &87.2 &91.1 &&98.9  &\textbf{92.7} &\textbf{95.7}  \\
			&InferCode   &93.2  &\textbf{92.6} &92.9 &&95.2 &90.3 &92.7  \\
			\hline
			\multirow{3}{*}{\textbf{Graph-based}} 	&SCDetector   &90.4  &87.1 &88.7 &&95.0 &89.2 &92.0  \\
			&FCCA   &\textbf{96.7} &89.8 &\textbf{93.1} &&94.1 &87.3 &90.6  \\
			&FA-AST   &96.3 &85.5 &90.6 &&94.7  &91.8 &93.2  \\
			\hline
			\multirow{1}{*}{\textbf{Our approach}} &\textbf{HELoC}  &\textbf{98.7}  &\textbf{95.9}  &\textbf{97.4}  &&\textbf{99.5} &\textbf{96.6}  &\textbf{98.0}     \\
			\toprule
		\end{tabular}
	}
		\vspace{-6pt}
\end{table}

\textbf{BCB.} Table 4 shows that all methods are very effective in terms of detecting Type-1 and Type-2 clones. 
InferCode uses the self-supervised framework by predicting subtrees to learn code representation and performs best in both recall and F1 scores compared to other baselines. However, it is still weaker than HELoC as it has the disadvantage of choosing TBCNN as the source code encoder. TBCNN captures the structural features of ASTs by sliding convolution kernels, which makes it difficult to capture long-term dependencies if the AST is deep or has many nodes. 
In particular, it treats ASTs as binary trees, which can change the original semantics of the source code, making long-term dependency problems even worse. 
In contrast, HELoC uses a specialized RSGNN to capture the global hierarchical structure and particularly allows the entire AST with path augmented node embeddings as input, alleviating the broken semantics.


Among the graph-based approaches, SCDetector and FCCA achieve good overall results as hybrid representation models for source code due to their mining of token, control flow and other information from the source code. However, their precision values are slightly lower for MT3 and Type-4 clones than the tree-based approaches.  For SCDetector, only tokens and CFGs are used, which have a significant advantage in reflecting the semantics of the code, but lacks the import of syntactic structures.
FCCA learns source code token, AST and CFG separately and then fuses them that achieves better than SCDetector. However, as a coarse-grained multi-view fusion method, it has the disadvantage that the semantic or syntactic structure of the code is not learned finely enough. Our approach adds textual content and position to the nodes and enables the fine-grained fusion of AST and code. In addition, our model demonstrates that learning the AST hierarchy is more likely to improve the detection precision of MT3 and Type-4 clone pairs. 

\textbf{GCJ and OJClone.} Table 5 shows that RtvNN achieves higher recall but very low precision. We find that RtvNN detects almost all the code pairs as clones. The reason could be that RtvNN relies on a simple distance metric to distinguish the hidden representation of each method generated, and does not handle the case where two functionally different methods may share syntactically similar components. SourcererCC achieves low recall and precision because it only considers the overlapping similarity of tokens between two code snippets and ignores the code snippet semantics, resulting in its inability to handle semantic clones.
Still, HELoC outperforms all baselines on both datasets
and achieves an F1 scores that is 4.3\% higher than the score obtained by FCCA on GCJ. 


\begin{table}[!htbp]
	\centering
	\caption{ Code clustering results on OJ and SA datasets.}
		\vspace{-6pt}
	\label{t1}
	\setlength{\tabcolsep}{3.8mm}{
		\begin{tabular}{c c c c }
		 \toprule 
			 \textbf{ Groups }  &\textbf{ Methods }    &\textbf{OJ}  &\textbf{SA}   \\ 
			\hline
			\multirow{2}{*}{\textbf{Token-based}} 
			&Word2vec         & 28.2    & 24.4     \\
			&Doc2vec       & 41.9    & 29.0 \\
				\hline
			 	\multirow{3}{*}{\textbf{Tree-based}} &Code2vec  & 58.2  & 51.4        \\
			 &Code2seq   & 52.9       & 49.0          \\
			 &InferCode   &\textbf{ 70.1}   &\textbf{ 62.0 }          \\
			 	\hline	
			 		\multirow{1}{*}{\textbf{Graph-based }} &GGNN       & 42.1    & 39.7        \\
		\hline
		\multirow{1}{*}{\textbf{Our approach}} &\textbf{HELoC} &\textbf{ 82.2}   & \textbf{74.6}        \\	
			\toprule
		\end{tabular}  
	}
	\vspace{-6pt}
\end{table}
\noindent\textbf{RQ3: How does our approach perform in code clustering?}
As shown in Table 6, our model can significantly improve the results of unsupervised clustering task. Overall, HELoC has a relative improvement of 12.1\% and 12.6\% in ARI metric on two datasets compared to the best competitor (Infercode).

The tree-based models significantly improve the token-based models. 
Code2vec and Code2seq model code snippets as a collection of paths in an AST, which captures the structural information of the code to some extent. However, the path representations they learned can only express fixed intervals between AST nodes at different levels rather than the dynamic distances learned by HELoC. This is why their results are not as good as our model. Infercode's decomposition of ASTs into subtrees breaks the overall semantic relationship of AST and leads to errors in understanding the overall structure of the code. HELoC aims to learn the topology of the AST autonomously without splitting the tree, which preserves the integrity of the code structure to a large extent.

\vspace{6pt}
\noindent\textbf{RQ4: What is the impact of different core components on the proposed model?  }
To evaluate the effectiveness of core components of HELoC, we conduct ablation experiments on three downstream tasks, and the experimental results are shown in Table 7. 

\emph{\textbf{No NEP.}} We investigate the impact of NEP by adding hierarchical information to the initial embedding of the nodes instead of automatically predicting it through pre-training. From the results shown in the table, it can be concluded that NEP impacts the model by learning the dynamic continuous distances of nodes at different levels.

\emph{\textbf{No NRO.}} To evaluate the impact of NRO, we remove it from HELoC. 
The results show that the model without NRO is the worst. 
NRO 
maps three topological relationships between node hierarchies to the embedding space. The experimental results confirm that introducing NRO 
can improve the performance of the model.

\emph{\textbf{No self-attention mechanism.}} We remove the self-attention component from HELoC. The results show that the model without this component performs less well on all datasets, indicating that the self-attention mechanism can improve the performance of the model by obtaining global information with a larger receptive field.

\emph{\textbf{No internal and external residual connections.}} To study the impact of residual connections, we compared HELoC with a variant obtained by removing the residual connections inside the self-attention mechanism. The results show that HELoC, including both residual connections inside and outside of the RSGNN, performs better since it effectively mitigates the problem of gradient vanishing inside the network.

{\emph{\textbf{No RSGNN.}} To evaluate the effectiveness of the feature encoder RSGNN, we replaced it with traditional GCN. From the overall results of the three downstream tasks, the loss of the model accuracy is significant after replacing RSGNN with GCN, 
demonstrating the critical role of RSGNN in HELoC.
}

\section{Threats To Validity}
\emph{\textbf{Threats to internal validity.}} They mainly come from the hyperparameters we set. 
Different hyperparameter settings affect the experimental results of baselines. We use the experimental results published in the original paper for baselines whom experimental data were provided. We retrain the baselines with the default parameters from the original paper to obtain the experimental results for those without provided data.

\emph{\textbf{Threats to external validity.}} They mainly come from the datasets we use. First, we conduct experiments on Java and C, so further work is still needed to validate our model to other programming languages. 
Second, several existing methods obtain results close to ours on the BCB dataset when performing code clone detection. In the future, we plan to use more challenging datasets with more functions 
to further evaluate the methods' capability. 


\section{Related Work}
\emph{\textbf{Source Code Representation.}} As stated in Section 1, how to represent source code effectively is a fundamental problem in software engineering research. Many source code representation techniques have achieved considerable success in many program comprehension tasks, including code classification \cite{liang2018automatic,lu2019program,mehrotra2020modeling}, malware detection \cite{pei2020amalnet,li2017software}, and code generation \cite{svyatkovskiy2020intellicode,sun2020treegen}. 
We have described major source code representations in Section 2.2 and pointed out their limitations in Section 2.3. 
Compared with these work, our proposed approach can better model the entire AST structures and achieve better performance in downstream tasks. 
\begin{table}[!htbp]
	\centering
	\caption{Effectiveness of core components of HELoC.}
		\vspace{-6pt}
	\label{t5}
	\setlength{\tabcolsep}{0.75mm}{
		\begin{threeparttable}
		\begin{tabular}{c c c c c c c c c c}
			\toprule
		\multirow{2}{*}{\textbf{Type}}   & \multicolumn{2}{c}{\textbf{Cla-}} &&	\multicolumn{3}{c}{\textbf{Clo- (F1)}}  &&\multicolumn{2}{c}{\textbf{Clu-}} \\
		\cmidrule(r){2-3} \cmidrule(r){4-7} \cmidrule(r){8-10} 
        	& GCJ  & OJ    && BCB-ALL	& GCJ  & OJClone   && OJ  & SA    \\ 
			\hline
	    \textbf{No\_NEP\tnote{1}}     & 88.3    & 86.1   && 83.6 & 89.2 & 87.7 && 70.1 & 63.4   \\
		\textbf{No\_NRO\tnote{2}}    & 75.1    & 72.4   && 70.7 & 68.1 & 73.4 && 60.1 & 49.2   \\
		\textbf{No\_SM\tnote{3}}     & 83.2    & 87.1   && 82.2 & 84.2 & 83.7 && 68.2 & 60.9   \\
  	    \textbf{No\_IERC\tnote{4}}    & 81.2    & 82.2   && 76.1 & 78.0 & 79.2 && 63.4 & 53.0   \\
		\textbf{No\_RSGNN\tnote{5}}     & 78.2    & 76.9   && 75.0 & 70.2 & 75.4 && 61.8 & 50.1   \\
		\textbf{HELoC}   &\textbf{97.2}    & \textbf{99.6}   && \textbf{98.0} & \textbf{97.4} & \textbf{98.0} && \textbf{82.2} & \textbf{74.6}   \\
		\hline
		\toprule
		\end{tabular}
	 \begin{tablenotes}
        \footnotesize
       \item[1] No node level prediction.
       \item[2] No node relationship optimization.
       \item[3] No self-attention mechanism.
       \item[4] No internal and external residual connections.
       \item[5] No residual self-attention GNN.
       
      \end{tablenotes}
  \end{threeparttable}

	}
	\vspace{-6pt}
\end{table}
Recently, several works \cite{hellendoorn2019global,shaw2018self,zugner2020language,peng2021integrating} inject structured information such as path embedding \cite{peng2021integrating} and node distance \cite{zugner2020language} into Transformer to introduce structural inductive bias. Unlike these Transformer-based models, we adopt a non-invasive strategy. Our RSGNN utilizes both graph convolutional networks (GCNs) (for capturing local structure) and the self-attention mechanism (for capturing global structure) to learn AST hierarchy comprehensively. In this way, our model allows a native AST as input without having to flatten it into the sequence required by Transformer, therefore the original structure of the code can be better preserved.

\emph{\textbf{Self-Supervised Learning.}} The advantage of self-supervised learning is that it benefits downstream tasks by designing pretext tasks to learn from large-scale unlabeled data. It has yielded considerable improvements in Natural Language Processing \cite{bui2021infercode,howard2018universal, devlin2018bert} and Computer Vision  \cite{erhan2010does, hao2019visualizing,gidaris2018unsupervised,pathak2016context}.
The contrastive learning approach \cite{hadsell2006dimensionality} has also received increasing attention as an important branch of self-supervised learning. It aims at minimizing the distance between similar data (positive) representations and maximizing the distance between dissimilar data (negative) representations. Promising results were obtained in the studies of He et al. \cite{he2020momentum} and Chen et al. \cite{chen2020simple}. Recently, such ideas have been embraced by researchers in the field of software engineering \cite{bui2021self, jain2020contrastive,chuang2020debiased,giorgi2020declutr,ding2021contrastive,chen2021varclr}. There are several choices of loss functions for contrastive learning, such as Triplet \cite{schroff2015facenet}, InfoNCE \cite{oord2018representation}, and Siamese \cite{chopra2005learning}.
In this paper, we design a novel hierarchical contrastive learning model to learn the relationships between nodes in an AST hierarchy.

\section{Conclusion}

\noindent In this paper, we propose HELoC, a hierarchical contrastive learning model for source code representation.
HELoC includes a novel hierarchical contrastive learning method to comprehensively capture the hierarchical features of ASTs by predicting the AST hierarchy through a dedicated RSGNN. 
To the best of our knowledge, HELoC is the first pre-training model that applies self-supervised contrastive learning for learning AST hierarchy.
We demonstrate the effectiveness of our approach by applying it to three downstream tasks, code classification, code clone detection, and code clustering, on multiple datasets. The experimental results show that our model outperforms other token-based, graph-based, and tree-based models.
The implementation of HELoC and the experimental data are publicly available at: \textbf{\url{https://github.com/Code-Rep/HELoC}}.

\begin{acks}
\noindent This work is financially supported by the Natural Science Foundation of Shandong Province, China (ZR2021MF059, ZR2019MF071), National Natural Science Foundation of China (61602286, 61976127) and Special Project on Innovative Methods (2020IM020100).
\end{acks}

\balance
\bibliography{sample-base}

\end{document}